\RequirePackage{luatex85} 
\documentclass{jfm}
\usepackage{dcolumn}% Align table columns on decimal point
\usepackage{supertabular,multirow}
\usepackage{epsfig}
\usepackage{bm}% bold math
\usepackage{subfigure}
\usepackage{color}
\usepackage{hyperref}% add hypertext capabilities
\usepackage{tikz,siunitx}
\usepackage{amssymb,amsfonts,amsmath}
\usepackage{euscript}
\usepackage{graphics}
\usepackage{setspace}
\usepackage{graphicx}
\usepackage{hyperref}% add hypertext capabilities
\usepackage{xcolor}
\usepackage{siunitx}
\usepackage[normalem]{ulem}
\usepackage[colorinlistoftodos]{todonotes}
\usepackage{soul}

\allowdisplaybreaks

\definecolor{brown}{rgb}{0.59, 0.29, 0.0}
\definecolor{orange}{RGB}{255,127,0}
\definecolor{brightube}{rgb}{0.82, 0.62, 0.91}

\graphicspath{{./},{Figs/},{Supply_Figs/}}

\definecolor{darkgreen}{rgb}{0, 0.5, 0.05}

\newcommand{\deleted}[1]{}

\newcommand{\eqn}[1]{\begin{align}#1\end{align}}
\newcommand{\bs}[1]{\boldsymbol{#1}}

\newcommand{\pare}[1]{\left( #1 \right) }

\newcommand{\fr}[2]{\frac{#1}{#2}}

\newcommand{\mc}[1]{\mathcal{#1}}

\newcommand{\tex}[1]{\mbox{\scriptsize{#1}}}

\def\dt{\Delta t}
\def\dd{\mathrm{d}}

\def\bbf{\bs{f}}

\def\bG{\bs{G}}

\def\bM{\bs{M}}

\def\bq{\bs{q}}

\def\br{\bs{r}}

\def\bu{\bs{u}}

\def\bv{\bs{v}}

\def\bzero{\bs{0}}
\def\bphi{\bs{\phi}}

\def\btau{\bs{\tau}}
\def\bomega{\bs{\omega}}

\def\blambda{\bs{\lambda}}

\def\mcB{\mc{B}}

\begin{document}

\title{Swimming Efficiently by Wrapping}

\author{H. Gidituri\aff{1},
  M. Ellero\aff{1,2,3}
  \and F. Balboa Usabiaga\aff{1} \corresp{\email{fbalboa@bcamath.org}}}

\affiliation{
  \aff{1} BCAM - Basque Center for Applied Mathematics, Alameda de Mazarredo 14, E48009 Bilbao, Basque Country - Spain
  \aff{2} Ikerbasque, Basque Foundation for Science, Calle de Maria Diaz de Haro 3, E48013 Bilbao, Basque Country - Spain
  \aff{3} Zienkiewicz Center for Computational Engineering (ZCCE), Swansea University, Bay Campus, Swansea SA1 8EN, UK
}

\date{\today}

\maketitle

\begin{abstract}
  Single flagellated bacteria are ubiquitous in nature.
  They exhibit various swimming modes using their flagella to explore complex surroundings such as soil and porous polymer networks.
  Some single-flagellated bacteria swim with two distinct modes,
  one with its flagellum extended away from its body and another with its flagellum wrapped around it.
  The wrapped mode has been observed when bacteria swim under tight confinements or in highly viscous polymeric melts. 
  In this study we investigate the hydrodynamics of these two modes inside a circular pipe.
  We find that the wrapped mode is slower than the extended mode in bulk but more efficient under strong confinement
  due to a hydrodynamic increase of its flagellum translation-rotation coupling
  and an Archimedes' screw-like configuration that helps to move the fluid along the pipe.
\end{abstract}

\section{Introduction}
\label{sec:intro}

Bacteria are prokaryotic microorganisms forced to live in a zero Reynolds number environment. 
Due to the kinematic reversibility of viscous flows, some bacteria have developed a non-reciprocal propulsion mechanism for locomotion, the rotation of flagella. 
The bacterium body and the flagella are rotated in opposite directions by molecular motors. 
Under rotation the flagella adopt an helical shape and propel the bacterium by working as a screw. 
Some bacteria can move both forward or backwards, in a push or pull mode, depending on the direction of rotation of the molecular motors 
and on the chirality of their flagella. 
As bacteria are often found in confined environments they have developed different strategies to swim while foraging in those conditions. 
One example is a swimming mode used by some monotrichous and bipolar bacteria 
where bacteria wrap their flagella around their own bodies resembling an Archimedes' screw (\cite{Kuehn2017, Tian2022, Thormann2022}). 
These bacteria swim alternating between two different modes, the wrapped mode and the extended mode, where in the latter the flagella extend away from their bodies.

The wrapped mode emerges when a bacterium encounter highly viscous or strongly confined environments (\cite{Kuehn2017}).
When a bacterium gets trapped during its forward pushing mode
a buckling instability occurs in the flagellar hook that triggers the flagellum wrapped mode (\cite{Kuehn2017, Park2022}).
The number of known bacterial species showcasing a wrapped mode under confinement is growing (\cite{Thormann2022}).
Thus, a natural question arises: is the wrapped mode a mere accident or is it selected due to some advantage to the bacteria?
Some studies suggest that the wrapped mode confer advantages to the motion in confinement environments.
Kühn et al. observed experimentally that the wrapped mode can enhance the motion in highly viscous and structured environments (\cite{Kuehn2018}).
Kinosita et al. studied the motion of bacteria with wrapped mode in very tight confinements and concluded that the wrapped mode
can allow the bacteria to glide over the substrate (\cite{Kinosita2018}).
Along this line of work we investigate how the flagella motion in the wrapped mode favors the motion of bacteria under strong confinement by hydrodynamic interactions only.
To this end we investigate the swimming of a bacterium inside circular pipes by means of CFD simulations.
We model the bacterium as a rigid ellipsoidal body with a single rigid flagellum attached to one of its pole
as done in previous works (\cite{Higdon1979, Lauga2006}).
The hydrodynamic interactions between the bacterium and pipe are computed by solving the Stokes equations with the rigid multiblob method,
a regularized boundary integral method (\cite{Usabiaga2016, Usabiaga2022}).
We show that the extended mode is more efficient in bulk and wide pipes while the wrapped mode can be more efficient in tight pipes.
The flow fields around the bacterium revel that the wrapped mode works as an Archimedes' screw pulling the fluid along the pipe.
The scheme of the paper is the following,
in Sec.\ \ref{sec:methods} we describe our numerical method and present some validations,
then, we describe our results in Sec. \ref{sec:bacteria_in_pipe} and conclude in Sec.\ \ref{sec:conclusions}.

\section{Numerical Method}
\label{sec:methods} 
We model a monotrichous bacterium as a rigid ellipsoid with an helical flagellum attached to one of its poles.
The flagellum is also modeled as a rigid object since it reaches an equilibrium configuration during steady state swimming (\cite{Higdon1979, Das2018}).
The body and the flagellum are connected by inextensible links that allow the flagellum to rotate freely around its main axis
but otherwise it is forced to move concomitant to the rigid ellipsoid.
Therefore, the bacterium's configuration is completely specified by seven degrees of freedom, the position and orientation of the body and
the flagellum rotation, or phase angle, around its axis (\cite{Higdon1979, Pimponi2016}).
The rigid objects, $\mcB_n$, move with linear and angular velocities, $\bu_n$ and $\bomega_n$, where we use the subindex $n$ to denote
either the bacterium body or the flagellum.
Due to the small bacterium size, the flow Reynolds number is vanishingly small, $\text{Re} \sim 10^{-5}$.
Thus, the flow can be modeled with the Stokes equations
\begin{align}
\label{eq:Stokeseqn}
- \nabla p + \eta \nabla^2 {\bs{v}} = \bzero, \\
\nabla \cdot {\bs{v}} = 0,
\end{align}
where $p$ and $\bv$ are the fluid pressure and velocity and $\eta$ its viscosity.
A no-slip boundary condition is imposed on the surface of the bacterium body and its flagellum
\begin{equation}
  \label{eq:no-slip}
  \bv(\br) = {\bs{u}_n} + {\bs{\omega}_n}\times ({\bs{r}}-{\bs {q_n}} )  \;\text{ for } \br \text{ on the bacterium,}
\end{equation}
where $\bq_n$ is tracking point of the rigid bodies (e.g.\ the bacterium body center and the flagellum attaching point respectively).

\begin{figure}
  \begin{center}                                    
    \includegraphics[width=0.9\columnwidth]{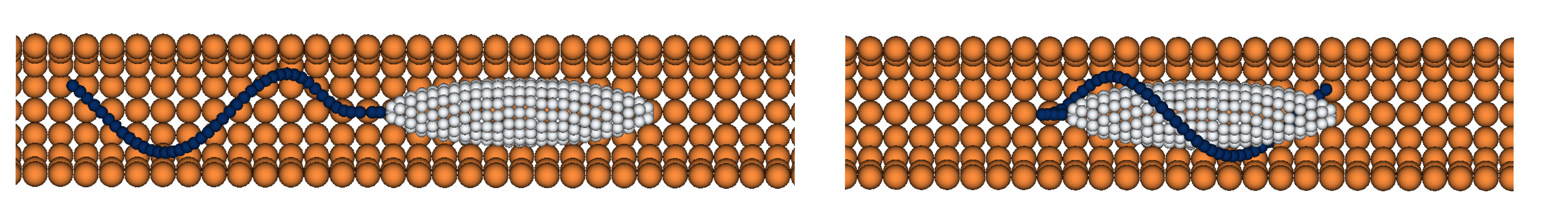} 
    \caption{Bacterium, $2.04\,\si{\mu m}$ long and $0.49\,\si{\mu m}$ wide, inside a pipe with inner radius $r_{0} = 0.4125\,\si{\mu m}$ with its flagellum in the
      extended and wrapped configuration.
      In both cases the flagellum has the same amplitude $\alpha = 0.32\,\si{\mu m}$ and the same number of waves along its axis $N_{\lambda} = 1.1$.
      Here we show a dissected section of the pipe to make the bacterium visible.}
    \label{fig:geometry}
  \end{center}
\end{figure}

To solve the coupled fluid-structure interaction problem  we use the rigid multiblob method for articulated bodies.
We summarized the numerical method while a detailed description can be found elsewhere (\cite{Usabiaga2022}).
The rigid bodies are discretized with a finite number of \emph{blobs} with position $\br_i$ as shown in Fig.\ \ref{fig:geometry}.
As the inertia is negligible the conservation of momentum reduces to the balance of force and torque.
The discrete force and torque balance for the rigid object $n$ can be written as,
\begin{align}
  \label{eq:force-balance-discrete}
  \sum_{i\in \mcB_n} \blambda_{i} - \sum_{i\in \mathcal{L}_n} \bphi_{n}  &= \bbf_n, \\
  \label{eq:torque-balance-discrete}
  \sum_{i\in \mathcal{B}_n} (\bs{r}_i  -{\bs {q}}_n) \times {\bs{\lambda}}_i - \sum_{i\in \mathcal{L}_n} (\Delta \bs{l}_{np}  -{\bs {q}}_n) \times {\bs{\phi}}_n  &= {\bs{\tau}}_n,
\end{align}
where $\bbf_n$ and $\btau_n$ are the external forces and torques acting on the rigid objects
while $\blambda_i$ are the constrained forces acting on the blobs that ensure the rigid motion of the bacterium body and the flagellum.
The second sums in \eqref{eq:force-balance-discrete}-\eqref{eq:torque-balance-discrete} run over the links, $\mathcal{L}_n$,
attached to the rigid object $n$ and $\bphi_{n}$ is the force exerted by the link $n$
to keep the rigid bodies connected while $|\Delta \bs{l}_{np}|$ is the link length.

The discrete no-slip condition evaluated at each blob $i$ is,
\begin{equation}
  \label{eq:no-slip-discrete}
  \bv(\br_i) = \sum_{j} \bM_{ij} \blambda_j =  {\bs{u}_n} + {\bs{\omega}_n}\times ({\bs{r}_{i}}-{\bs {q}_n} )  \; \text{ for } i\in \mathcal{B}_n.
\end{equation}
The mobility matrix $\bM_{ij}$ gives the hydrodynamic interaction between any two blobs, $i$ and $j$, of radius $a_i$ and $a_j$.
We use a regularized version of the Oseen tensor, the Rotne-Prager tensor (\cite{Wajnryb2013})
\begin{equation}
  \label{eq:rpy}
  \bM_{ij} = \fr{1}{(4\pi a_i a_j)^2} \int \delta (|{\bs{r'-r}_{i}}|- a_i) {\bs{G({\bs{r',r''}}})}\delta (|{\bs{r''-r}_{j}}|- a_j) \, \dd^3r' \dd^3r'' ,
\end{equation}
where $\bG(\br,\br')$ is the Green's function of the Stokes equations and $\delta(r)$ the Dirac's delta function.
The advantage of this formulation is that the regularized mobility has no divergence even when blobs get close
and it is not necessary to use special quadrature rules.
The equations \eqref{eq:force-balance-discrete}-\eqref{eq:no-slip-discrete} form a linear system for the unknown velocities, $\bu_n$ and $\bomega_n$,
and constraint forces, $\blambda_j$ and $\bphi_n$, that can be solved efficiently with iterative methods  such as GMRES (\cite{Usabiaga2016, Usabiaga2022}).

\subsection{Validation}
\label{sec:validation}

\begin{figure}
  \begin{center}
    \includegraphics[scale = 0.22]{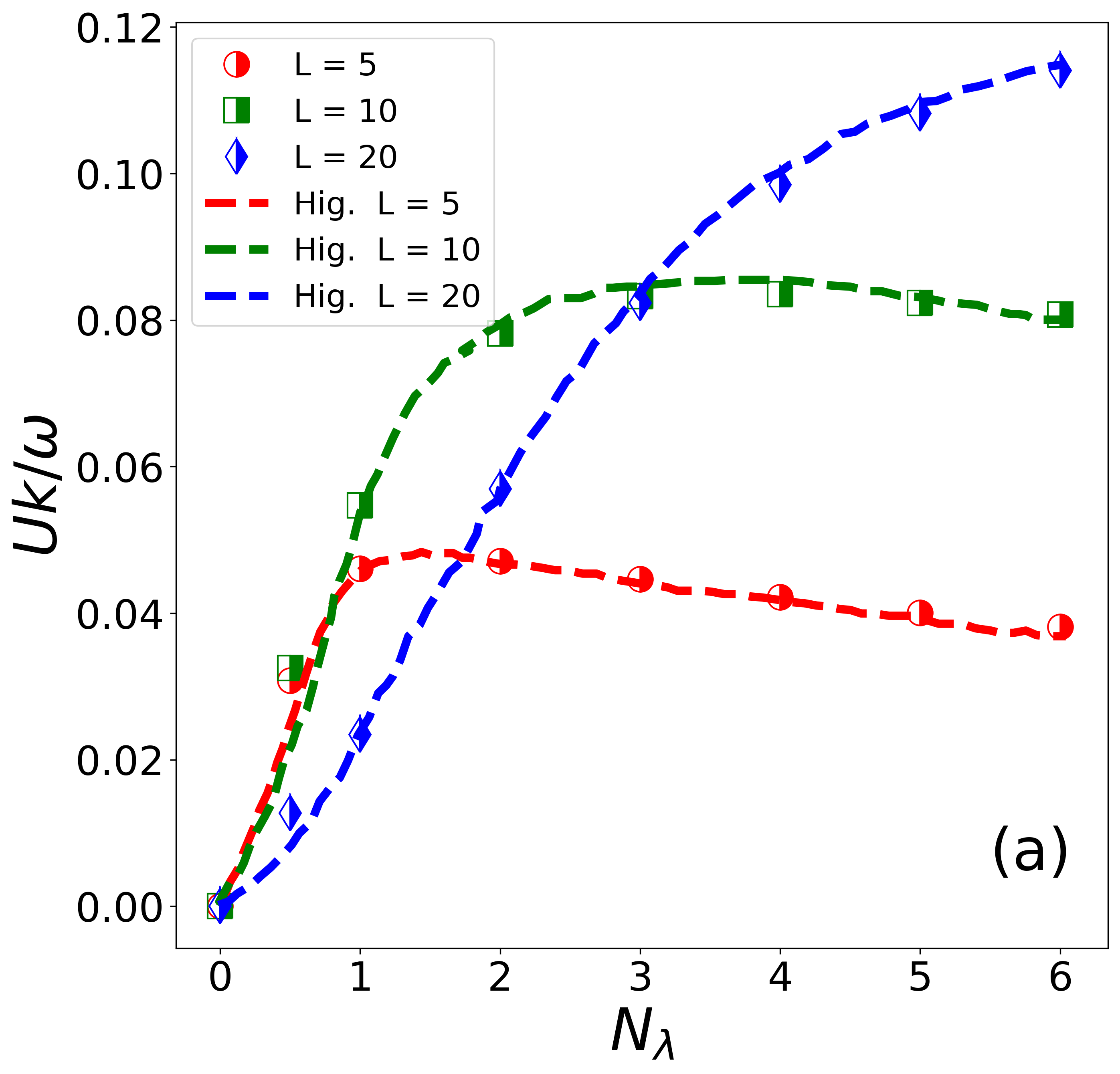}
    \includegraphics[scale = 0.22]{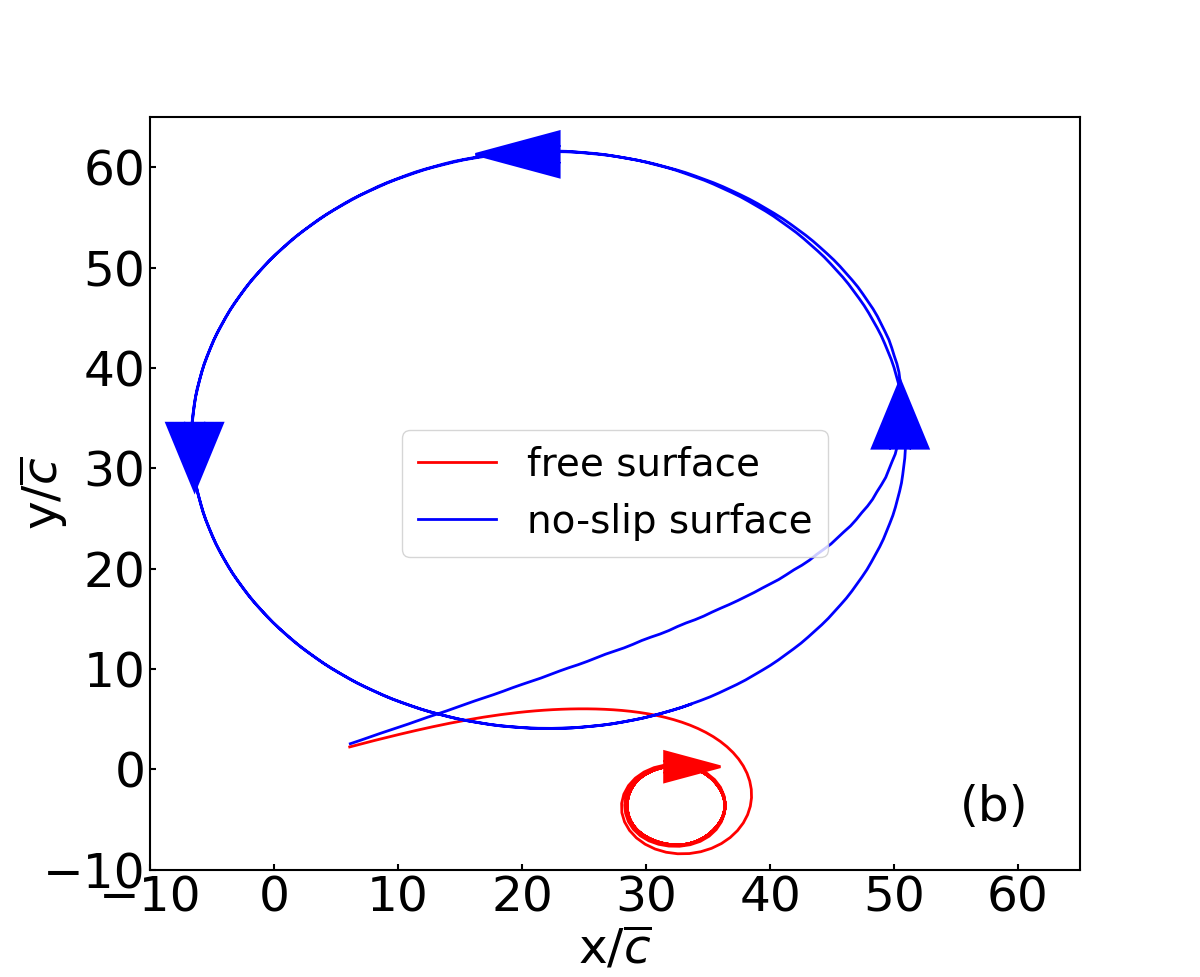}
    \caption{
      {\bf (a)} Non-dimensional swimming speed, $Uk/\omega_{\tex{motor}}$, plotted against the flagella number of waves, $N_{\lambda}=z_{\tex{max}}/k$,
      for bacteria with a spherical body of radius $c=1$ and different flagella lengths $L$.
      The flagella thickness, $\rho / c=0.02$, amplitude $\alpha$ and wavenumber $k=1 / \alpha$ are set to compare with the results of \cite{Higdon1979}.
      Dashed lines represent Higdon results while symbols represent our simulation results.
      {\bf (b)}  Circular trajectory comparison between a bacterium swimming above a no-slip surface and free surface.
      The arrows on the curves represent the direction of swimming.
    }
    \label{fig:U_bacteria_vs_N_lambda}
  \end{center}
\end{figure}

Here, we test our computational model against previous theoretical and computational works on helical flagellar swimming.
As first validation, we compute the swimming speed, in bulk, of a bacterium formed by a spherical body of radius $c=1$ and an helical flagellum of length $L$ and compare
our results against those of \cite{Higdon1979}.
As Higdon, we use flagella of helical shape with wavenumber $k$, amplitude $\alpha=1/k$ and thickness radius $\rho / c = 0.02$.
We discretize the bacterium body with $162$ blobs and the flagella with $125$, $250$ and $500$ blobs for the flagella lengths, $L = 5, 10, 20$ respectively.
We present in Fig.\ \ref{fig:U_bacteria_vs_N_lambda}(a)
the dimensionless velocity,  $U k / \omega$, where $\omega$ is the relative angular velocity between the bacterium body and flagellum versus
the number of waves in the flagellum, $N_{\lambda}$.
The agreement with the results of Higdon is good for all the flagella considered.

As second validation we compute the trajectory of a bacterium above a no-slip and a free-slip surface.
The equations of motion are integrated with a midpoint scheme which requires solving two mobility problems per time step to find the bacterium body and
flagellum's velocities (\cite{Usabiaga2022}).
The bacterium body is modelled as an ellipsoid with major axis length $4$ and aspect ratio $\gamma = 2$, discretized with $287$ blobs.
The flagellum of length $L=11.91$ and $N_{\lambda} = 2$ is discretized with $154$ blobs.
We also incorporate a short-range steric repulsion between the blobs and the surface to avoid possible overlaps.
We use a steric repulsion of strength  $f=0.01 \, \si{pN}$ for overlapping blobs with the surface
and with an exponential decay with a characteristic length $\xi=0.01\,\si{\mu m}$ for non-overlapping blobs.
It is well known from previous studies that flagellated bacteria move in circular trajectories,
of opposite directions,
when swimming close to a no-slip or a free-slip surface (\cite{Lauga2006, Shum2010, Pimponi2016}).
In addition, \cite{Pimponi2016} reported that the radius of the trajectory is small in the free-slip case in comparison with no-slip case.
Our results captures the previously observed features, the bacterium is hydrodynamically attracted to the surface and swim in circles,
see Fig.\ \ref{fig:U_bacteria_vs_N_lambda}(b).

\section{Results and Discussion}
\label{sec:bacteria_in_pipe}

In this section we study the swimming of a bacterium inside circular pipes of radius $r_0$ and length $L_0 \approx 21 r_0$ aligned along $z$.
Keeping the aspect ratio constant ensures that the flow disturbance created by the bacterium decays to negligible values at the pipes ends (\cite{Liron1978}).
We model the pipes as immobile rigid objects discretized with blobs just as the bacterium (\cite{Usabiaga2016}).
We place the bacterium in the middle of the pipes and we use that configuration to compute the bacterium velocity.
Since the Stokes equations assume steady state flows, the velocity in a given configuration can be obtained by solving the mobility problem one single time.
Later, we will consider the case where the bacterium freely swims inside a periodic pipe. 

We consider two different swimming modes.
First, the extended mode where the flagellum is attached to the body front part and it extends away from it.
In the second mode the flagellum is wrapped around the bacterium body, see Fig.\ \ref{fig:geometry}.
In both cases we apply constant and opposite torques, of magnitude $\tau_m=0.46\,\si{pN\mu m}$, to the body and the flagellum to model the work exerted by a molecular motor.
Thus, we assume that the molecular motor always works in the constant torque (low frequency) regime (\cite{Xing2006}).
In most numerical experiments the flagellum extends along its main axis a length similar to the bacterium body.
Thus, in the wrapped mode the body is fully covered by the flagellum.
The bacterium body, always $2.04\,\si{\mu m}$ long and $0.49\,\si{\mu m}$ wide, is discretized with 292 blobs of radius $a=0.0425\,\si{\mu m}$.
The geometric details of the helical flagella and pipes used in this work
are presented in Tables \ref{tab:flagella} and \ref{tab:pipe}.

\begin{figure}
  \begin{center}
    \includegraphics[scale = 0.2]{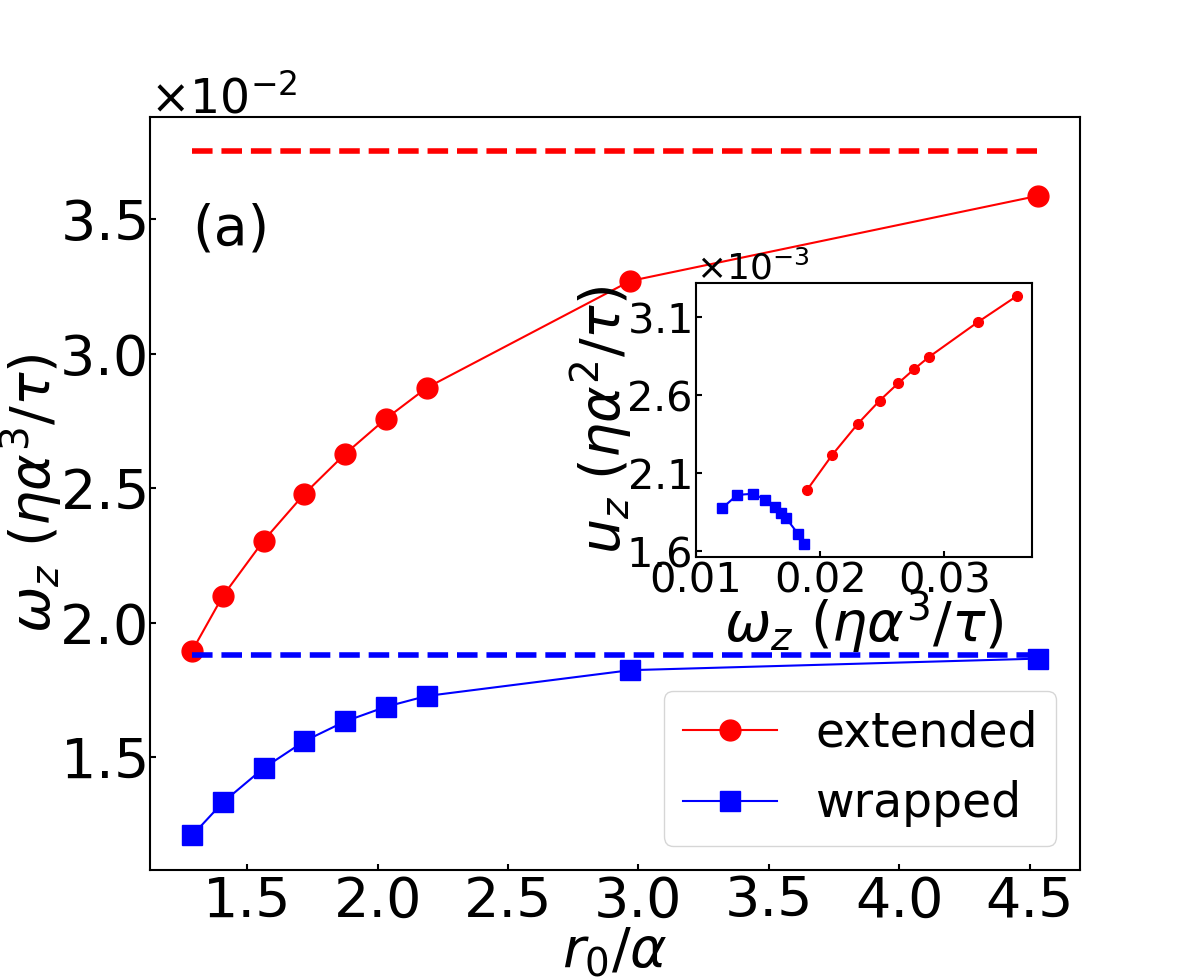}
    \includegraphics[scale = 0.2]{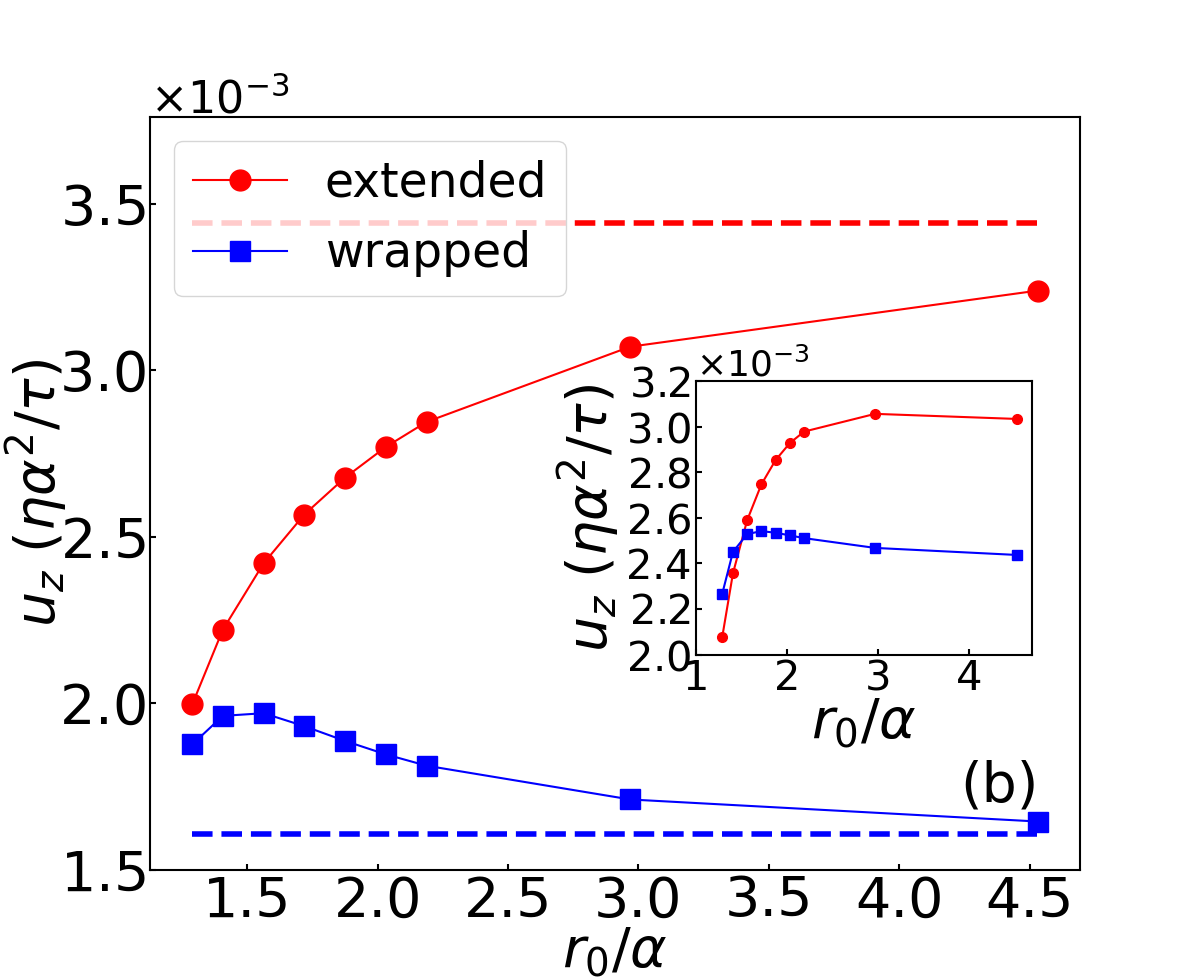}
    \caption{Flagellum angular velocity, $\omega_z$, {\bf (a)}
      and bacterium swimming speed along the pipe, $u_z$, {\bf (b)}  versus pipe radius $r_0$.
      Results for the extended and wrapped mode using the same flagellum (model II) with
      length $L=3.6,\si{\mu m}$ and amplitude $\alpha=0.32\,\si{\mu m}$.
      The dashed lines depict bulk values.
      The inset in {\bf (a)} shows the swimming speed against the flagellum angular frequency for all the pipe radius used.
      The inset in {\bf (b)} shows the swimming speed calculated with a minimal model, Eq.\ \ref{eq:swim_vel}.
    }
    \label{fig:speed_omega_extended_wrapped}
  \end{center}
\end{figure}

All the motion is driven by the rotation of the flagellum.
Therefore, we start looking at its angular velocity, $\omega_z$, see Fig.\ \ref{fig:speed_omega_extended_wrapped}a.
In bulk the flagellum rotates two times faster in the extended mode than in the wrapped mode.
The slower rotation can be explained by the additional drag experienced by the flagellum in the wrapped mode,
which is caused by the proximity of the flagellum to the bacterium body.
Both modes reduce their angular velocities as $r_0$ decreases due to the additional hydrodynamic drag generated by the pipe walls.
However, the decrease is proportionally less important in the wrapped mode as its initial drag was larger.
Thus, the ratio between the angular frequencies of the two modes falls from a factor $\sim 2$ in bulk to a factor $\sim 1.6$ in the smallest pipe considered.

Next, we look at the swimming speed along the pipe axis, $u_z$, see Fig.\ \ref{fig:speed_omega_extended_wrapped}b.
We observe that in bulk the wrapped mode swims about twice slower than the extended mode.
This result is consistent with experimental observations  (\cite{Kuehn2017, Tian2022, Grognot2021, Thormann2022}).
The slower swimming speed in the wrapped mode is a consequence of the slower rotation of its flagellum.
Under confinement the swimming speed, $u_z$, decreases for the extended mode as the pipe radius is decreased.
Again, the additional hydrodynamic drag generated by the pipe walls is responsible for this effect.
In contrast, the wrapped mode exhibits a non-monotonic trend in its swimming speed.
As the pipe radius is decreased the bacterium swims faster up to the point where the ratio between the pipe radius
and the flagellum amplitude is $r_0 / \alpha \approx 1.5$.
Beyond that point the swimming speed decreases with $r_0$.
The striking difference between both swimmer modes is also shown
in the inset of Fig.\ \ref{fig:speed_omega_extended_wrapped}a, which shows the swimming velocity versus the flagellum angular velocity for all
pipe radius.
Because the Stokes equations are linear the linear and angular velocity are proportional when keeping all geometric parameters constant.
However, changing the pipe radius affects the proportionality constant dramatically for the wrapped mode and only weakly for the extended mode.
A resolution study shows that these results are maintained after increasing the resolution of our discretization by a factor 4 and 16,
see Appendix \ref{sec:resolution_study} and Fig.\ \ref{fig:resolution}.

\begin{figure}
  \begin{center}
    \includegraphics[scale=0.2]{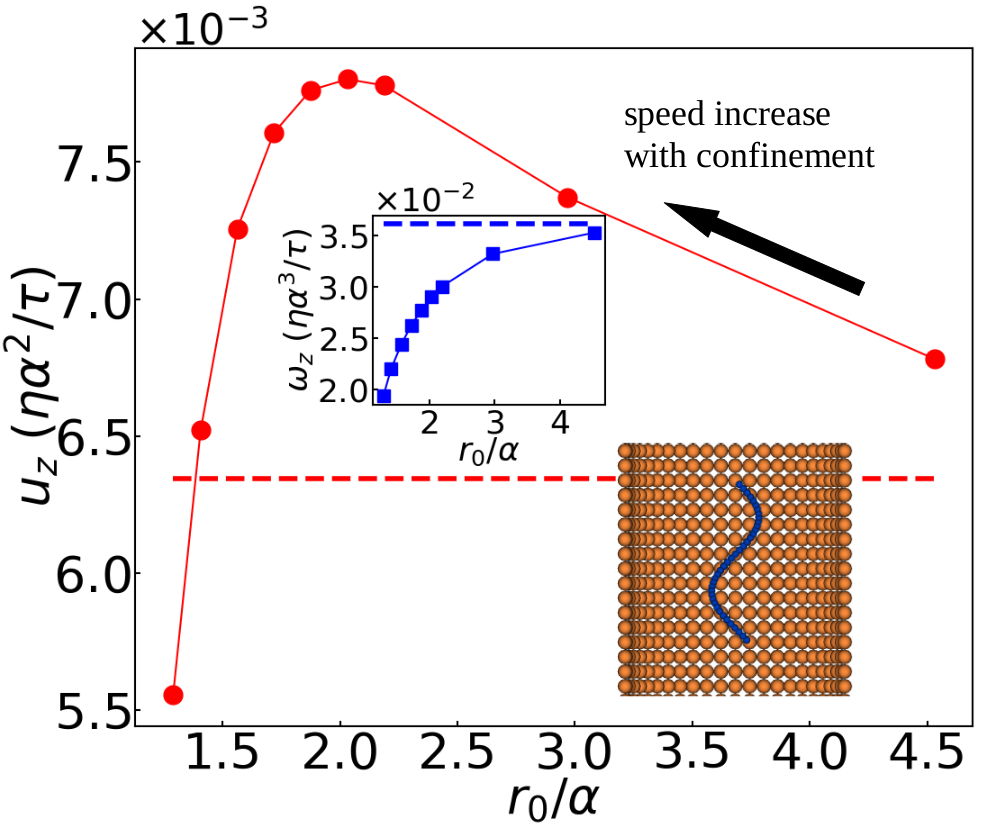}
    \caption{Swimming speed, $u_z$, and z-component of the angular velocity (inset), $\omega_z$, of a single helical flagellum 
      with amplitude $\alpha=0.32\,\si{\mu m}$ (model II) inside a pipe of radius $r_0$. 
      The dashed lines depicts the bulk values and a snapshot of the flagellum in a pipe is shown as inset. 
    } 
    \label{fig:helical_flagella} 
  \end{center} 
\end{figure}

We seek a physical mechanism that explains the difference between swimming modes and the speed enhancement in the wrapped case. 
We start by considering the motion of a single helical flagellum inside a pipe. 
We apply a constant torque on the helical flagellum and measure its translational and rotational speeds. 
Note that in this case the flagellum is not a torque-free swimmer, as there is no body to which apply an opposite torque. 
Nonetheless, this numerical experiment is useful to understand the more complex wrapped mode. 
We observe an increase in the swimming speed for decreasing pipe radius with respect to the bulk value above a critical pipe radius, see Fig.\ \ref{fig:helical_flagella}, 
similar to the wrapped mode results. 
For the single flagellum its swimming speed can be written as $u_z = M_{tr} \tau_z$. 
For moderate confinements the hydrodynamic interactions with the wall increase the value of the mobility coupling term, $M_{tr}$, with respect to the bulk values, 
thus, the swimming speed is increased. 
For very tight confinements the lubrication interactions dominate the interactions with the wall and $M_{tr}$ decreases below the bulk values. 
These effects were already reported by \cite{Liu2014a} for an infinite flagellum within an infinite pipe. 
This speed increase is observed despite the reduction in the flagellum angular velocity, $\omega_z$, with $r_0$, see Fig.\ \ref{fig:helical_flagella} inset.

For the whole bacterium the situation is more complex because
at the same time that the flagellum thrust is enhanced an additional drag on the bacterium body tends to reduce the swimming speed.
This interplay between the enhanced thrust and the additional drag has been observed in a recent experimental study with \emph{E. coli} (\cite{Vizsnyiczai2020}).
Vizsnyiczai et al.\ observed that a bacterium swimming in an extended mode inside a pipe swims slower than a bacterium in a channel.
However, when the bacterium is exiting the pipe and only its flagellum remain inside, the swimming speed is larger than in a channel showing 
the increased translation-rotational coupling experienced by the flagellum and the lack of an additional drag acting on the bacterium body (\cite{Vizsnyiczai2020}).
Our numerical results agree with the experimental work of Vizsnyiczai et al.\
for the extended mode while they suggest that for the wrapped mode the enhanced thrust is the dominant effect.

\begin{figure}
  \begin{center}
    \includegraphics[width=0.95 \columnwidth]{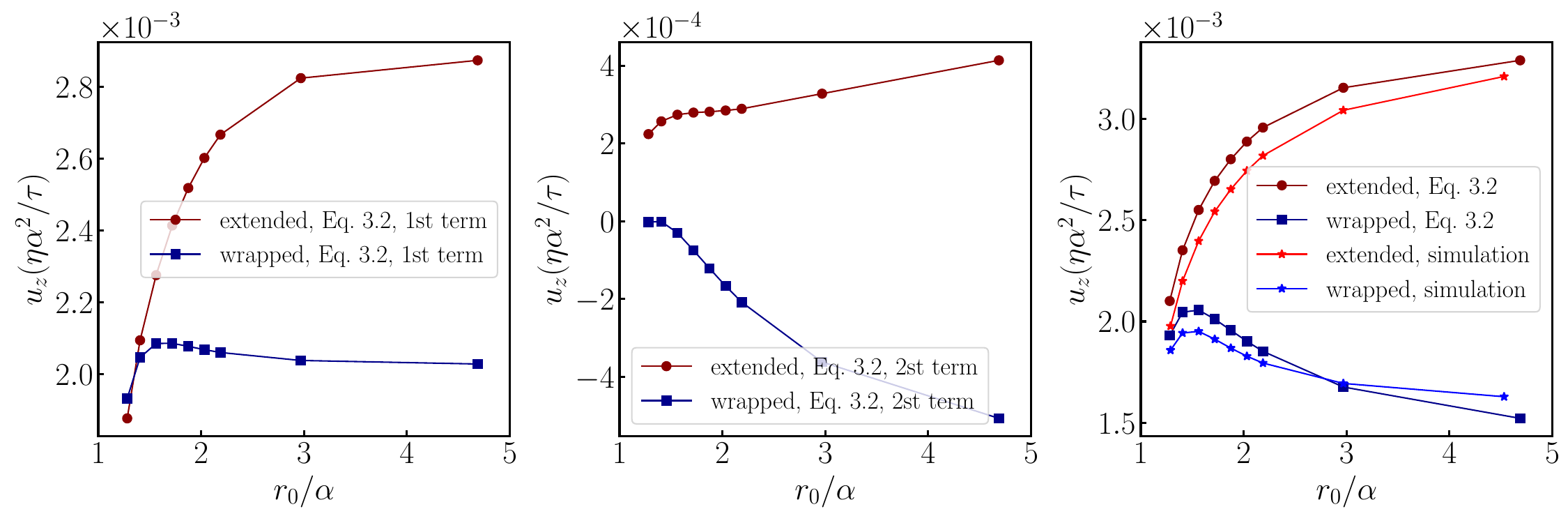}
    \caption{Swimming velocity normalized by the motor torque predicted by the first term and second terms of the minimal model,
      Eq.\ \ref{eq:swim_vel}, panels {\bf (a)} and {\bf (b)} respectively and
      the full Eq.\ \ref{eq:swim_vel} compared with the hydrodynamic simulations, panel {\bf (c)}.   
    }
    \label{fig:minimal_model}
  \end{center}
\end{figure}

% The increased flagellum translation-rotation coupling under confinement speed ups both swimming modes, however, it is only the dominant effect for the wrapped mode. 
To explore in more detail the difference between both swimming configurations we introduce a minimal model 
in which we treat the bacterium body and flagellum as two connected rigid bodies that move exactly along the pipe axis 
and are free to rotate along their axes (\cite{Vizsnyiczai2020}).
Thus, in the minimal model the velocities perpendicular to the pipe axis are assumed to be zero unlike in the previous simulations.
Taking into account that the bacterium is a free-force-torque swimmer we can write the linear system
\eqn{ 
  \label{eq:linear_system}
  \begin{bmatrix}
    u_z \\
    \omega^{\tex{body}}_z \\
    u_z \\
    \omega^{\tex{flag}}_z
  \end{bmatrix} =
  \begin{bmatrix}
    M_{tt} & 0 & C_{tt} & C_{tr} \\
    0 & M_{rr} & C_{rt} & C_{rr} \\
    C_{tt} & C_{rt} & N_{tt} & N_{tr} \\    
    C_{tr} & C_{rr} & N_{tr} & N_{rr} \\
  \end{bmatrix}
  \begin{bmatrix}
    f_z \\
    \tau_z \\    
    -f_z \\
    -\tau_z
  \end{bmatrix}.
}
All the mobility components in \eqref{eq:linear_system} depend on the pipe radius and the bacterium mode (extended or wrapped)
due to the hydrodynamic interactions.
The unknowns in the linear system are the swimming velocity, $u_z$, the angular velocities of body and flagellum,
$\omega^{\tex{body}}_{z}$ and $\omega^{\tex{flag}}_{z}$ and the force acting on each of them $f_z$.
Solving these equations for the swimming velocity we obtain
\begin{align}
  \label{eq:swim_vel}
  u_z = -\frac{\pare{M_{tt} - C_{tt}}\pare{N_{tr} - C_{rt}}}{M_{tt} + N_{tt} - 2C_{tt}} \tau_z +
  \frac{\pare{N_{tt} - C_{tt}} C_{tr}}{M_{tt} + N_{tt} - 2C_{tt}} \tau_z. 
\end{align}
We compute numerically all the mobility components and plot the contribution of the first and second terms of Eq.\ \eqref{eq:swim_vel} in Fig.\ \ref{fig:minimal_model}.
The panel (a) shows that the first term in \eqref{eq:swim_vel} is the dominant one.
It also shows that the decay in the speed is more pronounced for the extended mode which occurs
due to a stronger decay of $M_{tt}$ and $N_{tr}$ due to the hydrodynamic drag generated by the pipe.
The contribution of the second term of \eqref{eq:swim_vel} is qualitatively different for both modes, as shown in Fig.\ \ref{fig:minimal_model}b.
The mobility component $C_{tr}$ decays to almost zero for the wrapped mode but not for the extended mode.
For the wrapped mode that term was a negative contribution, thus, when it vanishes it leads to an overall speed increase.
The minimal model, including both the first and second term of \eqref{eq:swim_vel}, explains the full numerical result reasonably well as shown in in Fig.\ \ref{fig:minimal_model}c.
Thus, the main difference between the two swimming modes is lower decrease of the coefficients $M_{tt}$ and $N_{tr}$ in the wrapped mode.

\begin{figure}
  \begin{center}                                    
    \includegraphics[width=0.95 \columnwidth]{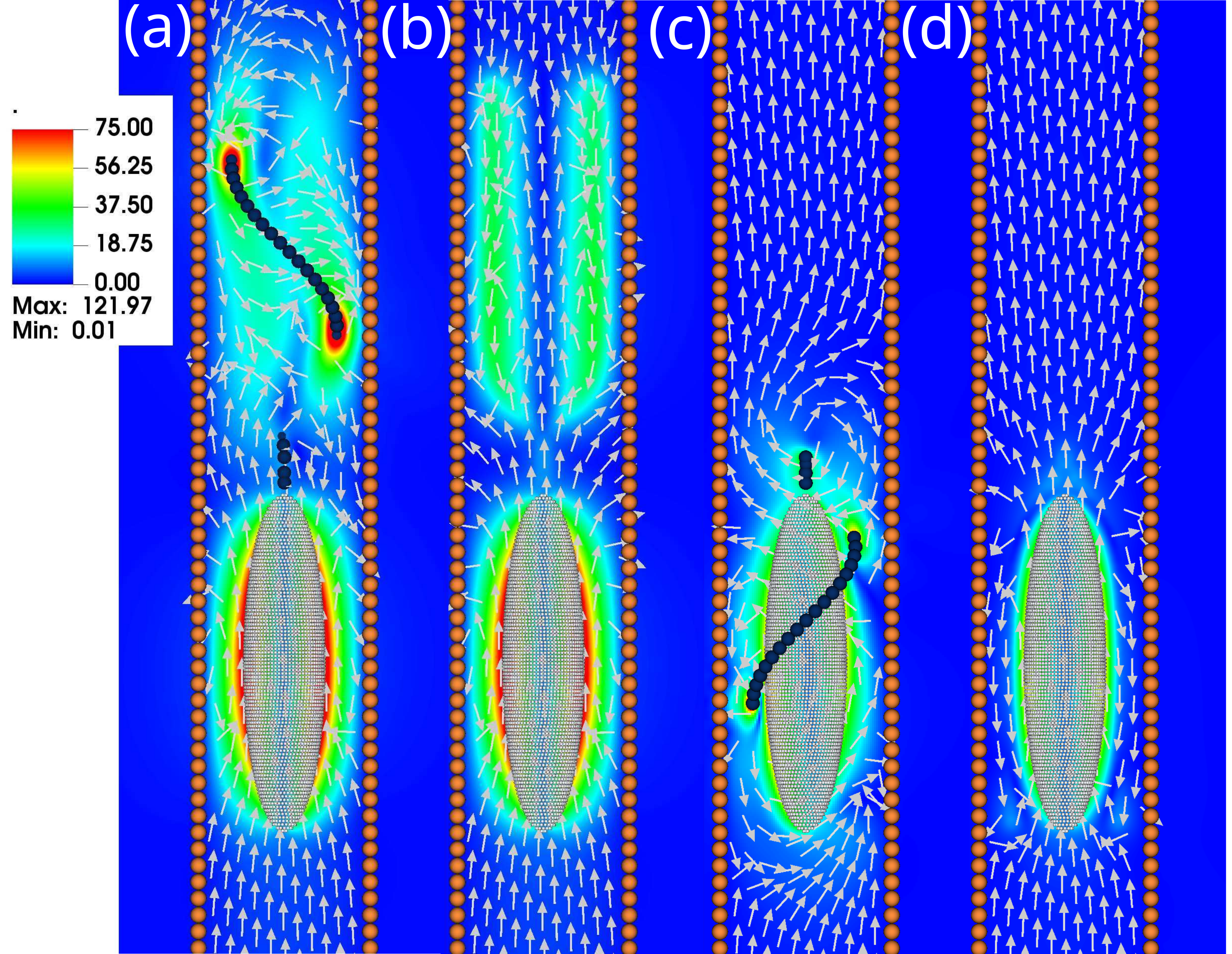} 
    \caption{
      In plane velocity fields around a bacterium inside a pipe with inner radius $r_0/\alpha = 1.56$.
      Instantaneous and average flow field during one flagella rotation for the extended mode, {\bf (a)} and {\bf (b)} respectively,
      and for the wrapped mode {\bf (c)} and {\bf (d)}.
      The flagellum is not shown in the panels with the average flows ({\bf (b)}, {\bf (d)}).
      See also Movie 1.
    }
    \label{fig:flows}
  \end{center}
\end{figure}

To further understand the difference between both swimming modes we compute the flow around the bacterium in a tight pipe, with $r_0/\alpha = 1.56$,
using the highest resolution model, with $\sim 16$ number of blobs than the original simulations, see Appendix \ref{sec:resolution_study}.
For the extended mode we show the instantaneous and the average flow field during one flagellum rotation in Fig.\ \ref{fig:flows}(a,b).
The same flows for the wrapped mode are shown in Fig.\ \ref{fig:flows}(c,d).
As the bacterium moves along the pipe an equivalent volume of fluid is displaced to make room for the bacterium,
after all, the fluid is incompressible.
The fluid can be pushed further along the pipe or transported to the back of the bacterium
but, interestingly, flow perturbations decay exponentially fast along the pipe (\cite{Liron1978}),
therefore, it follows that in long enough pipes all the fluid displaced is moved to the back of the bacterium.
We can see in Fig.\ \ref{fig:flows}(b) a counter flow close to the pipe wall.
However, the high velocity gradients suggest a large drag that we observed as a reduction in the bacterium swimming speed.
In the wrapped mode the flagellum works as an Archimedes' screw helping to push fluid to the back of the bacterium, see Fig.\ \ref{fig:flows}(d).
The effect of the flagellum pushing the fluid is observed more clearly in the supplementary Movie 1.
This is the ultimate mechanism that differentiates both swimming modes and makes
the additional drag on the body the dominant effect for the extended mode and the enhanced rotation translation coupling the dominant one for the wrapped mode.

\begin{table}
  \begin{center}
    \begin{tabular}{ccccccccc}
      Model & $L \,[\si{\mu m}]$ & $\alpha \,[\si{\mu m}]$ & $k\, [\si{\mu m^{-1}}]$ & $L_z \,[\si{\mu m}]$ & $N_{\lambda}$ & $a \,[\si{\mu m}]$ & $N_b$ \\
      \hline
      I   & 3.29   & 0.32  & 2.27   & 2.57   & 0.8  & 0.0425  & 40 \\
      II  & 3.68   & 0.32  & 3.13   & 2.40   & 1.1  & 0.0425  & 43 \\
     % II & 3.5372 & 0.32  & 2.8280 & 2.5681 & 1.0  & 0.0425  & 45 \\
      III & 4.27   & 0.32  & 4.25   & 2.57   & 1.5  & 0.0425  & 52 \\
      IV  & 5.11   & 0.32  & 5.66   & 2.57   & 2.0  & 0.0425  & 60 \\
      V   & 6.00   & 0.32  & 7.06   & 2.57   & 2.5  & 0.0425  & 72 \\
      VI  & 5.17   & 0.32  & 2.83   & 3.84   & 2.83 & 0.0425  & 69 \\
%     VII & 3.6783 & 0.32  & 3.1250 & 2.3982 & 1.1  & 0.0425  & 43 \\
    \end{tabular}
    \caption{Flagella model parameters.
      Flagellum length, $L$, amplitude, $\alpha$, wave number, $k$, 
      maximum extension along its axis, $L_z$,
      number of waves along its axis, $N_{\lambda}=L_{z} / \lambda = L_z k / (2\pi)$, blob radius, $a$, and number of blobs $N_b$.
      The amplitude of the wave was exponentially damped near the attaching point with a damping factor $k_E=k$ as in Ref.\ \cite{Higdon1979}.
    }
    \label{tab:flagella}
  \end{center}
\end{table}

\begin{table}
  \begin{center}
    \begin{tabular}[t]{cccc}
      Model & $L \,[\si{\mu m}]$ & $r_0 \,[\si{\mu m}]$ & $N_b$  \\
      \hline
      I    & 9.35 & 0.4125 & 720  \\
      II   & 10.1 & 0.45   & 816  \\
      III  & 11.1 & 0.50   & 1008  \\
      IV   & 12.1 & 0.55   & 1159  \\
      V    & 13.1 & 0.60   & 1386  \\
    \end{tabular}
    \begin{tabular}[t]{cccc}
      Model & $L \,[\si{\mu m}]$ & $r_0 \,[\si{\mu m}]$ & $N_b$  \\
      \hline
      VI   & 14.1 & 0.65   & 1562  \\
      VII  & 15.1 & 0.7    & 1824  \\
      VIII & 20.1 & 0.95   & 3232  \\
      IX   & 30.1 & 1.45   & 7248  \\
    \end{tabular}    
  \end{center}
  \caption{Pipes model dimensions.
    Length, $L$, inner radius $r_0$ and number of blobs $N_b$.
    The blob radius is $a=0.1\,\si{\mu m}$ in all cases.
  }
  \label{tab:pipe}
\end{table}

\subsection{Power and Efficiency}
\label{sec:Power_Efficiency}

\begin{figure}
  \begin{center}                                    
    \includegraphics[width=0.45\columnwidth]{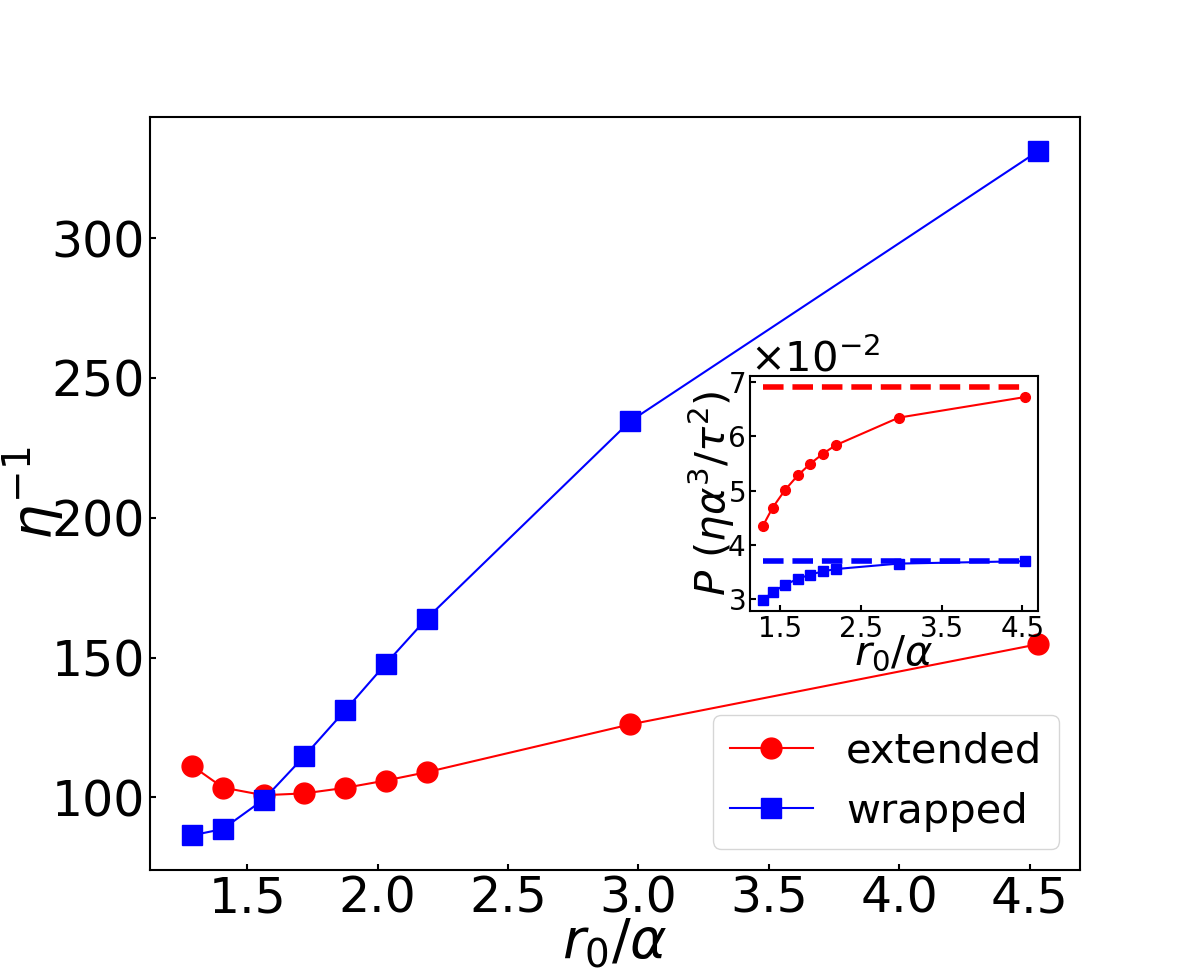}
    \caption{Inverse of the efficiency versus radius of the pipe ($r_0$).
      The inset figure shows the swimming power and the dashed lines depict the bulk values.
    }
    \label{fig:power}
  \end{center}
\end{figure}

The power consumption is an important quantity for a microswimmer propelling in a viscous environment
and the efficiency can be more important than the absolute swimming speed.
Thus, we measure these quantities.
Considering the chemical energy used within the cell is beyond the scope of our work, thus, we limit ourselves to study the power dissipated by the Stokes flow
and the microswimmers hydrodynamic efficiency.
The power exerted by a microswimmer to the medium and dissipated by the flow is
\eqn{
  \label{eq:P}
  P = \sum_n \pare{\bbf_n \cdot \bu_n + \btau_n \cdot \bomega_n},
}
where the sum is over rigid bodies, in our case the bacterium body and its flagellum. 
As the power is generated by the motor, the power consumed by a bacterium during its swimming can be rewritten as
$P_m = \btau_m \cdot \bomega_m = \btau_m \cdot (\bomega_{\text{flag}} - \bomega_{\text{body}})$.
In the absence of elastic or soft steric interactions both expressions are equivalent.
We will always use \eqref{eq:P} to account for the soft steric interactions used in Sec. \ref{sec:dynamics}.

The wrapped mode consume less power for all pipe radii owing to the slower rotation of its flagellum, see Fig.\ \ref{fig:power} inset.
Under confinement the power exerted by the motor decays for both swimming modes.
Of more interest is the hydrodynamic efficiency of the swimmers to propel themselves.
There are several approaches to define the hydrodynamic efficiency (\cite{Childress2012}).
We follow a classical approach and define the inverse efficiency as the power normalized with the
power necessary to pull the bacterium body with the same speed (\cite{Higdon1979, Lauga2013})
\eqn{
  \eta^{-1} = \fr{M_{zz}}{u_{z}^2} P,
}
where $M_{zz}=u_{z}/f_{z}$ is body mobility (without flagellum) along the pipe axis and $u_z$ the velocity.
Figure \ref{fig:power} shows the inverse efficiency as a function of the pipe radius.
In bulk and wide pipes the extended mode is more efficient.
However, there is a crossover and for tight confinements the wrapped mode becomes more efficient.
This is a result of the lower power consumption of the wrapped mode and, importantly, its enhanced velocity within the pipe.
This result suggests that the wrapped mode has an advantage to propel in confined spaces.
To verify the accuracy of the results we perform a resolution study by increasing the resolution by a factor 4 and 16 and we obtain essentially the same
results as shown in Appendix \ref{sec:resolution_study} and the Fig.\ \ref{fig:resolution}.
So far we have only used one flagellum, model II, and a bacterium placed exactly on the middle of the pipe.
In the next two sections we explore whether these results are robust under a change in these conditions.

\subsection{Robustness of results: Effect of $N_{\lambda}$ and L}
\label{sec:effect_of_N_lambda}

\begin{figure}
  \begin{center} 
    \includegraphics[width=0.27\columnwidth]{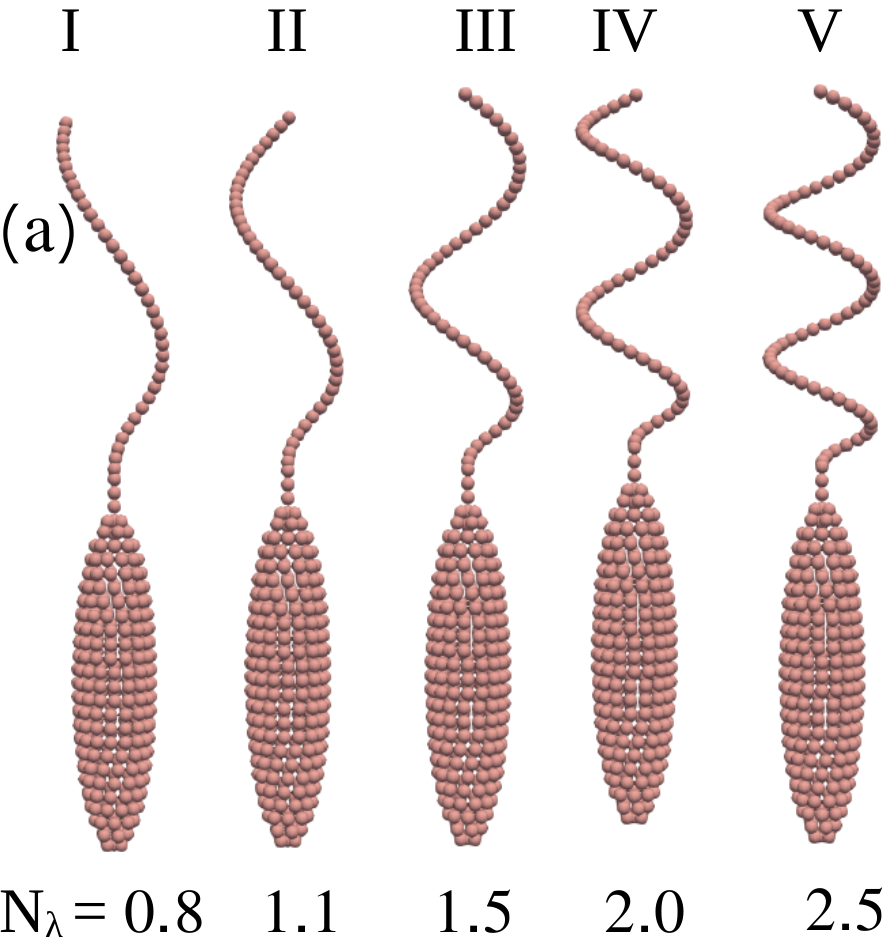}
    \includegraphics[width=0.3375\columnwidth]{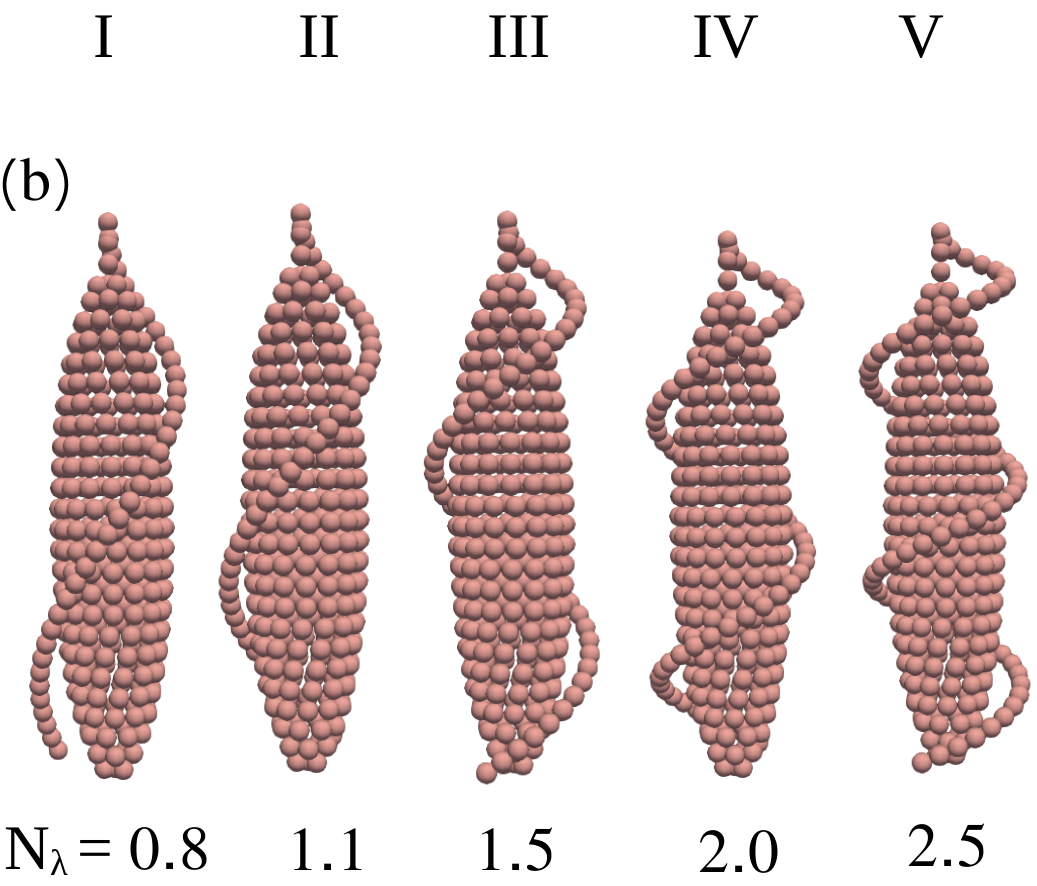}
    \includegraphics[width=0.42\columnwidth]{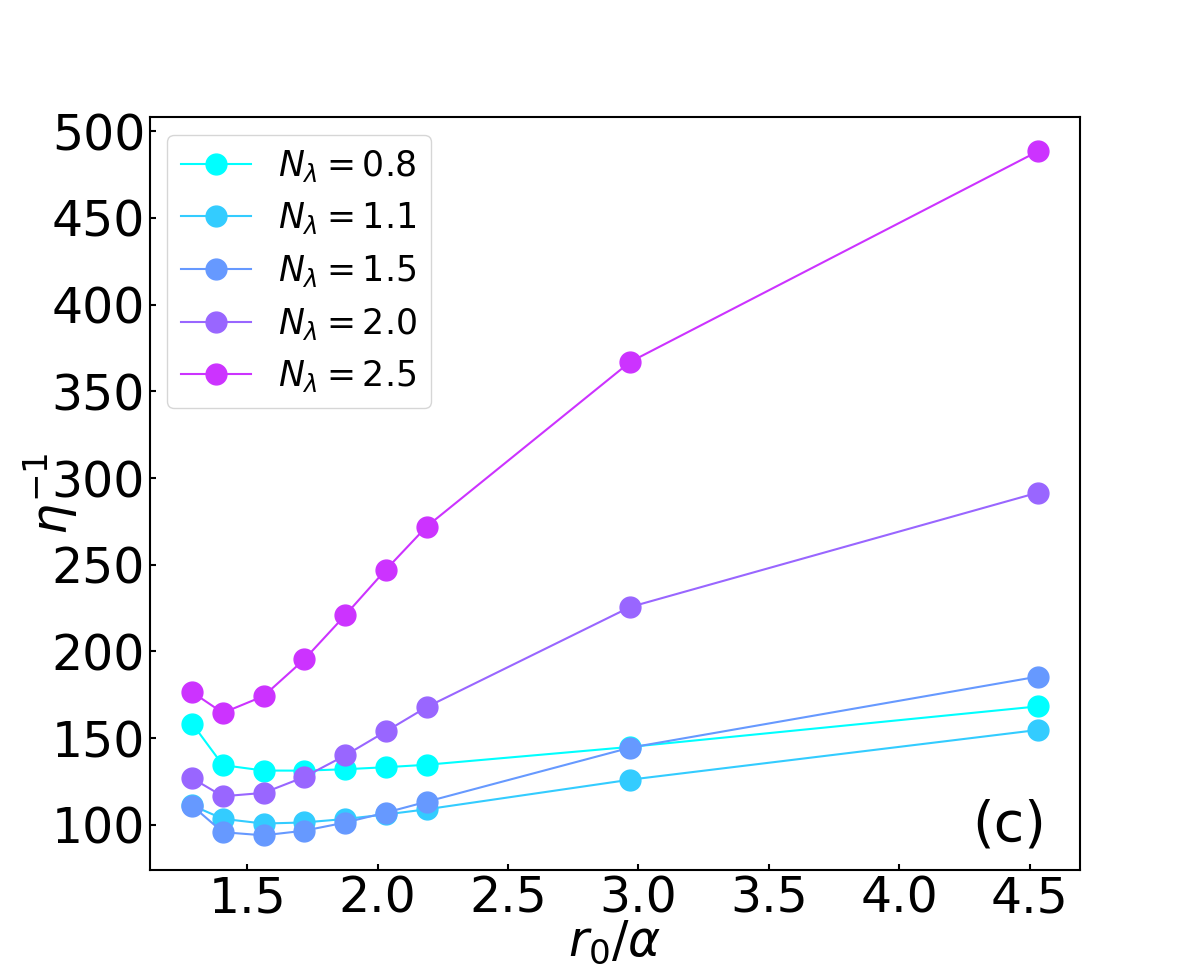}
    \includegraphics[width=0.42\columnwidth]{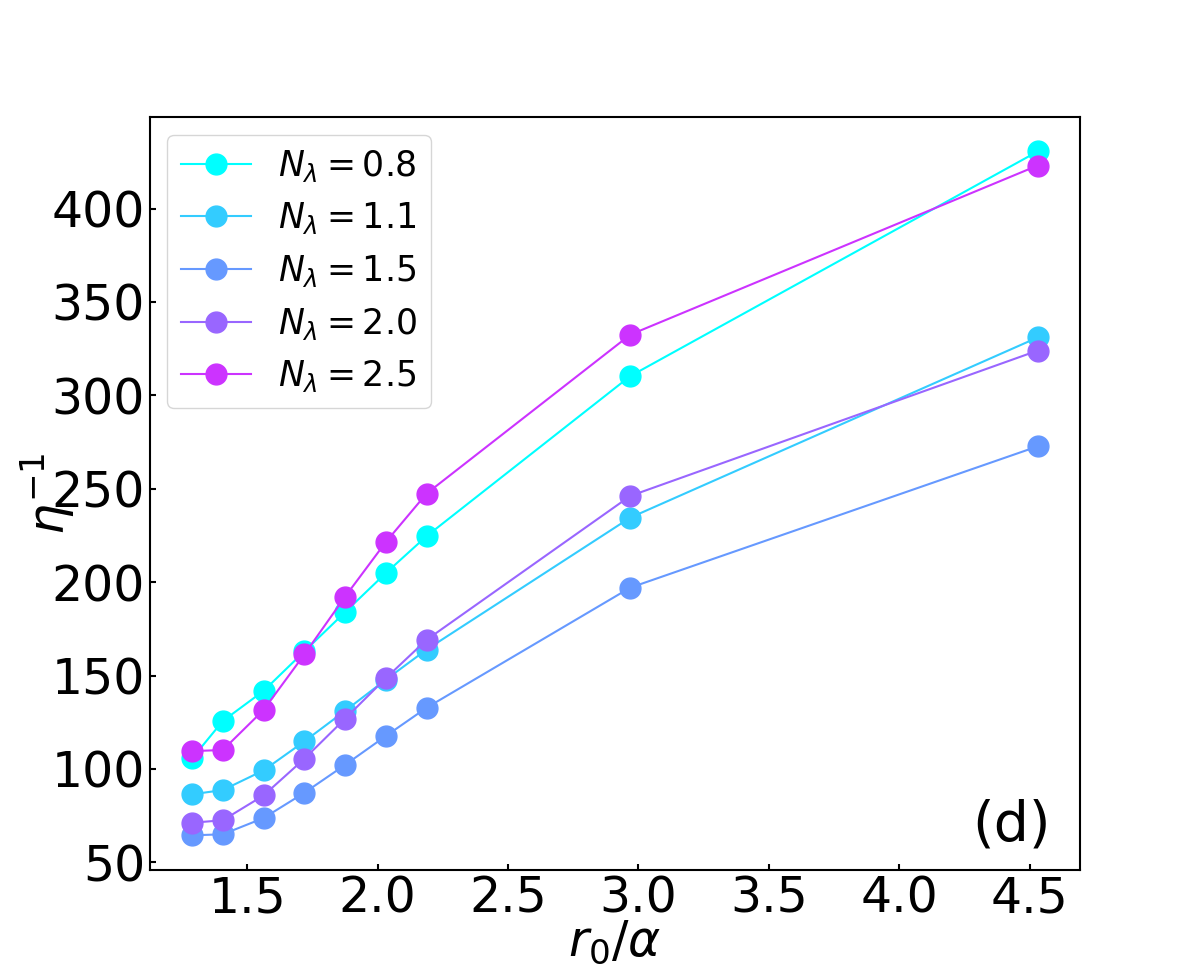}
    \caption{Shapes of extended {\bf (a)} and wrapped {\bf (b)}  bacteria modes with different $N_{\lambda}$ values.
      Inverse of the swimming efficiency versus radius of the pipe ($r_0$) for different values of $N_{\lambda}$
      for the extended mode {\bf (c)} and the wrapped mode {\bf (d)}.
    }
    \label{fig:geometry_N_lambda}
  \end{center}
\end{figure}

Bacterial species present flagella of different lengths, amplitudes and pitch angles
which affect the bacteria bulk speeds and efficiencies (\cite{Higdon1979, Lauga2013}).
Here, we explore if the wrapped mode is a more efficient swimming style in confined environments for a wide variety of flagella models.
We build five flagella models by varying simultaneously the flagellum length, $L$, and the number of waves along its length,
$N_{\lambda}=L_z / \lambda$, where $L_z$ is the flagellum extension along its axis and $\lambda$ the
wavelength of the helical wave, see Fig. \ \ref{fig:geometry_N_lambda}(a,b) and Table \ref{tab:flagella}.
We present the inverse efficiency for all flagella models and pipe radius in Fig.\ \ref{fig:geometry_N_lambda}(c,d)

The general trend is the same as before.
For wide pipes the extended mode is more efficient than the wrapped mode for all flagella models except one ($N_{\lambda}=2.5$).
Under confinement both swimmers increase their efficiency but the improvement is stronger for the wrapped mode which becomes the most efficient 
for pipes with $r_0 / \alpha \lessapprox 1.7$.
In those situations the wrapped mode is approximately two times more efficient than the extended mode.
The efficiency, for both swimming modes, is non-monotonous on $N_{\lambda}$.
When $N_{\lambda} \ll 1$ the flagellum is almost straight, thus, it cannot propel the bacterium.
Therefore, the swimming speed and the efficiency initially grow with $N_{\lambda}$.
Beyond a certain value of $N_{\lambda}$ the flagellum tangent forms a large angle with the direction of motion, which again reduces the propulsion efficiency.
For intermediate values of $N_{\lambda}$ the flagellum is helical-shaped which allows propulsion.
For both modes the flagellum with $N_{\lambda}=1.5$ is the most efficient under confinement for the flagella lengths considered.
For bacteria swimming in bulk the optimum is also close to $N_{\lambda}=1.5$, although the exact optimum $N_{\lambda}$ depends on the flagellum length (\cite{Higdon1979}).
For the extended mode, optimal swimming occurs around the non-dimensional pipe radius, $r_{0}/\alpha$ = 1.5 for all values of $N_{\lambda}$.
For the wrapped mode the optimal swimming occurs for lower values of $r_{0}/\alpha$.

\subsection{Robustness of results: dynamical simulations}
\label{sec:dynamics}

So far we have computed the swimming speed when the bacterium is located in the middle of the pipe and aligned along it.
However, freely swimming bacteria can tilt and move towards the pipe wall.
To verify if the results reported so far are robust, we perform dynamic simulations where the bacterium is free to displace away
from the pipe centerline and to change orientations.
We use the same pipe models as before but imposing  periodic boundary conditions along the pipe.
To solve the Stokes equations with these boundary conditions we use a periodic Fast Multipole Method implemented in the library STKFMM (\cite{Yan2018a}).
To avoid the overlap of the bacterium with the pipe we include
a steric repulsion interaction between the blobs of pipe and bacterium with a repulsion strength of $f=0.05 \, \si{pN}$
for overlapping blobs and with an exponential decay with a characteristic length $\xi=0.01\,\si{\mu m}$ for non-overlapping blobs.
As in Sec.\ \ref{sec:validation} we use a midpoint integrator to integrate the equations of motion (\cite{Usabiaga2022}).
In all simulations we use the same time step size, $\dt=10^{-3}\, \si{s}$, so a full flagella rotation takes around 12 (36) time steps for the fasted (slowest) rotating flagella.
For all models considered in this section we simulate the bacterium for $10\,\si{s}$ so the bacterium can swim at least $70\,\si{\mu m}$.
We use the last $8\,\si{s}$ to extract the swimming speed and the power consumption.

\begin{figure} 
  \begin{center}
    \includegraphics[width=0.95 \columnwidth]{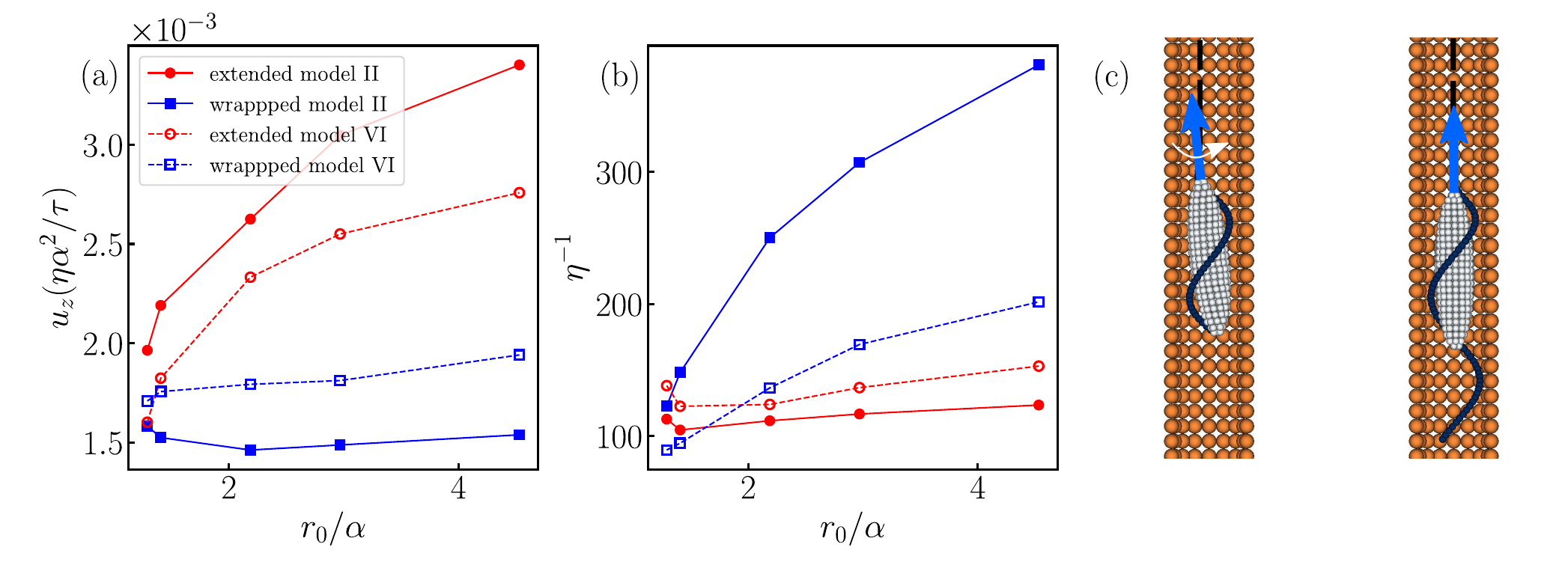}
    \caption{Swimming speed {\bf (a)} and hydrodynamic efficiency {\bf (b)} of wrapped and extended modes swimming along a periodic pipe
      extracted from $10\,\si{s}$ long dynamic simulations.
      Full symbols use flagellum model II, which extends the same length as the body, and open symbols use model flagellum VI, which extends longer.
      The longer flagellum prevents the tilt of the wrapped mode which results in higher speeds and better efficiencies.
      {\bf (c)} Two bacteria inside a pipe.
      A bacterium with a short flagellum performs a precession motion around the pipe axis while a bacterium with a long flagellum is forced to swim straight.
      The swimming direction is shown with a large arrow and the pipe axis with a dashed line.
    }
    \label{fig:dynamics}
  \end{center}
\end{figure}

We observe that the bacterium swimming in the extended mode swims near the pipe walls while bacterium swimming in the wrapped mode tries to remain away from them.
This lateral displacement is more evident in wide pipes where the bacterium has more space to move, see Movie 2,
and it is reminiscent of the hydrodynamic attraction to walls by the extended mode mentioned in Sec.\ \ref{sec:validation}
and the hydrodynamic repulsion from walls experienced by the wrapped mode (not shown).
The swimming speed and efficiency, for the bacterium with the flagella model II, are shown in Fig.\ \ref{fig:dynamics}.
The same general trend as for the static simulations is observed.
However, the efficiency curves do not cross over.
The cross over is not observed because this time the wrapped swimming speed along the pipe, $u_z$, barely increases with confinement,
and the efficiency depends strongly on $u_z$.
The magnitude of $u_z$ does not increase because the bacterium swims with a tilt towards the wall, see Fig.\ \ref{fig:dynamics}c and Movie 3.
In contrast, the extended mode cannot tilt significantly on small pipes as that is prevented by its rigid flagellum, which favours the motion along the pipe.

To verify if the whole extension of the swimmers affects the result we run another set of simulations using a longer flagellum, model VI,
that extends beyond the bacterium body, see Fig.\ \ref{fig:dynamics}c and Movie 4.
The results are presented as open symbols in Fig.\ \ref{fig:dynamics}.
In this case the speed of the wrapped mode is approximately independent on the confinement but larger than with the shorter flagellum.
As a result we observe a crossover between the efficiencies of the wrapped and extended modes.
Overall, these results show that
\emph{(i)} the swimming speed is less sensitive to confinement for the wrapped mode than for the extended mode,
\emph{(ii)}, the efficiency improves strongly for the wrapped mode and
\emph{(iii)}, depending on the flagellum details, the wrapped mode can be the most efficient way to swim under confinement.

\section{Conclusions}
\label{sec:conclusions}
In this paper we have presented the dynamics of two different swimming modes, namely the extended and wrapped modes of monotrichous type bacteria.
Under bulk conditions the extended mode swims faster and more efficiently than the wrapped mode.
However, under strong confinement the efficiency of the wrapped mode improves faster than for the extended mode.
For a wide number of flagella shapes, with different lengths and wavelengths, the bacterium in the wrapped mode swims more efficiently.
For both swimming modes the flagellum rotation translation coupling is enhanced by the pipe confinement (\cite{Liu2014a, Vizsnyiczai2020}).
However, the drag on the bacterium body is also increased and only for the wrapped mode the enhanced flagellum coupling is the dominant effect.
The flow fields around the bacterium suggest that the wrapped mode works as an Archimedes' screw helping to transport the fluid displaced by
the motion of the bacterium.
Such screw-like configuration improves the swimming efficiency.

These results are complementary to the experimental work of Kinosita et al.
where the bacteria \emph{Burkholderia} adopting the wrapped mode was observed to glide in very narrow ducts (\cite{Kinosita2018}).
It seems that, either by gliding over a substrate or by means of hydrodynamic interactions, the wrapped mode promotes the motion
of bacteria on tight confinements.
It is interesting to note that some bipolar flagellated bacteria can display a wrapped and an extended mode simultaneously,
where the flagellum at the front pole wraps around the body and the rear one remains extended (\cite{Murat2015, Constantino2018, Thormann2022, Bansil2023}).
Such mixed mode could present some advantages under confinement that should be investigated.

\vspace{0.25cm}
\noindent
{\bf Declaration of Interests:} The authors report no conflict of interest.

\section*{Acknowledgments}
The project that gave rise to these results received the support of a fellowship from ``la Caixa''
Foundation (ID 100010434), fellowship LCF/BQ/PI20/11760014, and from the European Union's Horizon 2020 research and innovation
programme under the Marie Skłodowska-Curie grant agreement No 847648. 
Funding provided by the Basque Government through the BERC 2022-2025 program
and by the Ministry of Science and Innovation: BCAM Severo Ochoa accreditation
CEX2021-001142-S/MICIN/AEI/10.13039/501100011033 and the project PID2020-117080RB-C55
``Microscopic foundations of soft matter experiments: computational nano-hydrodynamics (Compu-Nano-Hydro)'' are also acknowledged.

\appendix
% \section{Appendix-I}
\section{Resolution study}
\label{sec:resolution_study}

\begin{figure}
    \centering
    \includegraphics[width=0.45 \columnwidth]{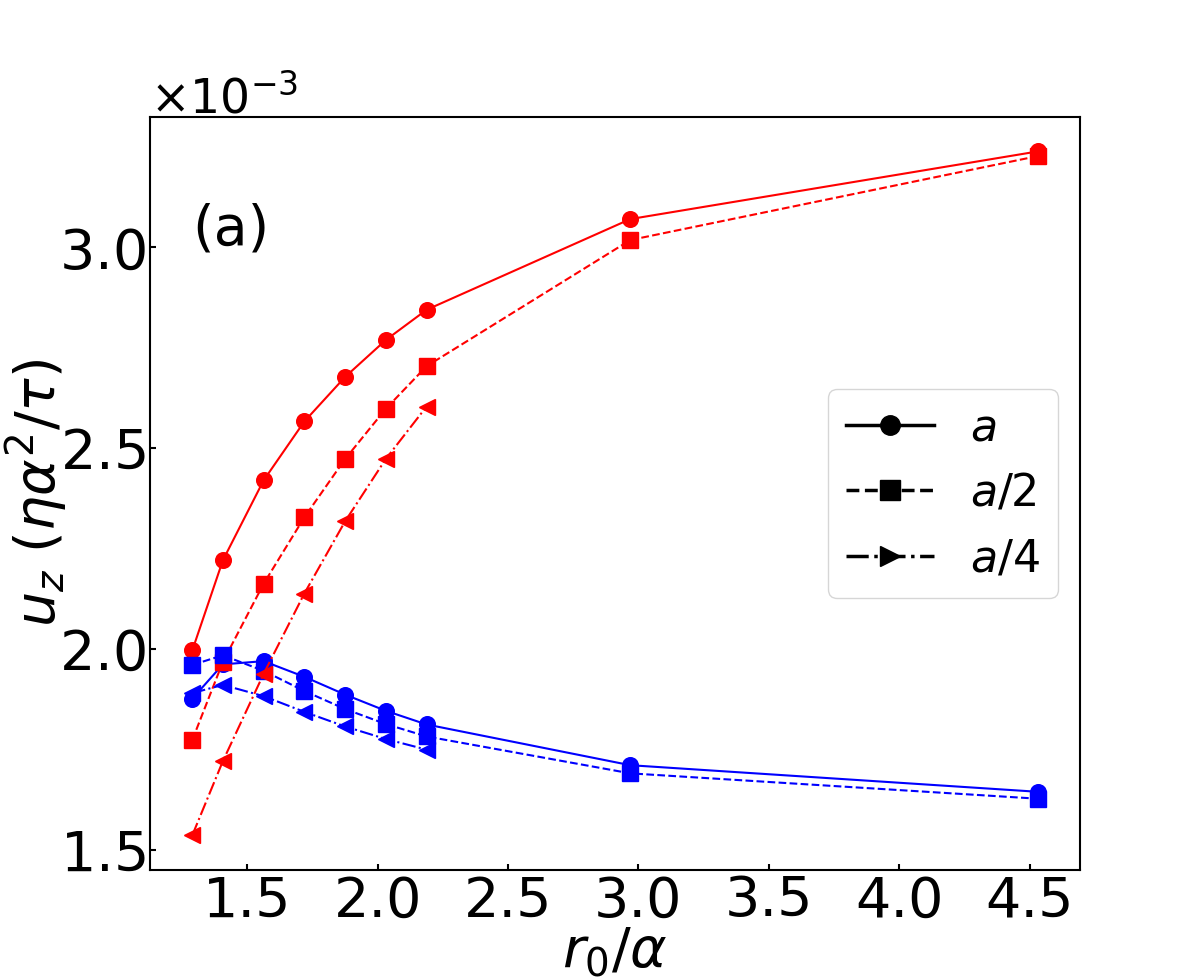}
    \includegraphics[width=0.45 \columnwidth]{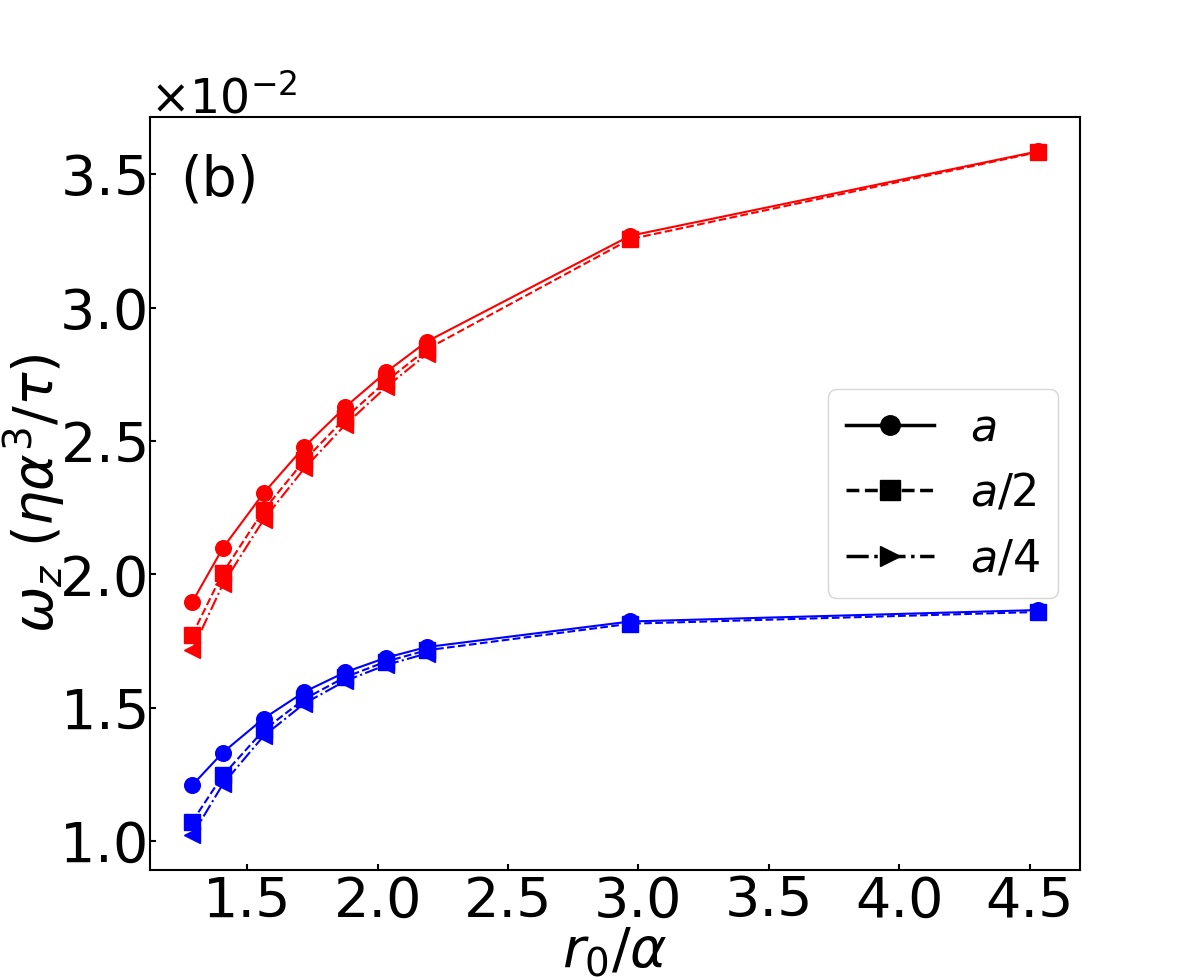}
    \centering
    \includegraphics[width=0.45 \columnwidth]{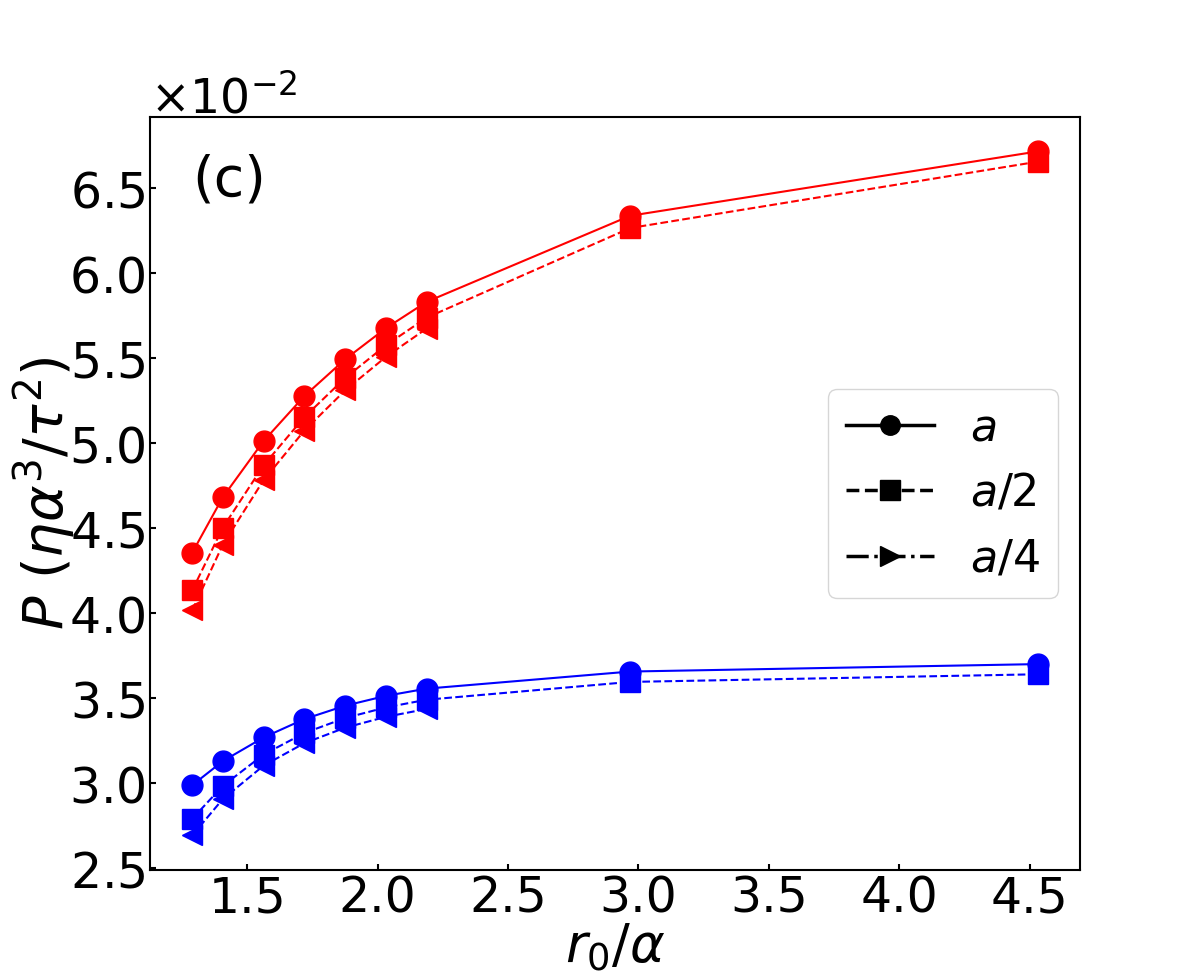}
    \includegraphics[width=0.45 \columnwidth]{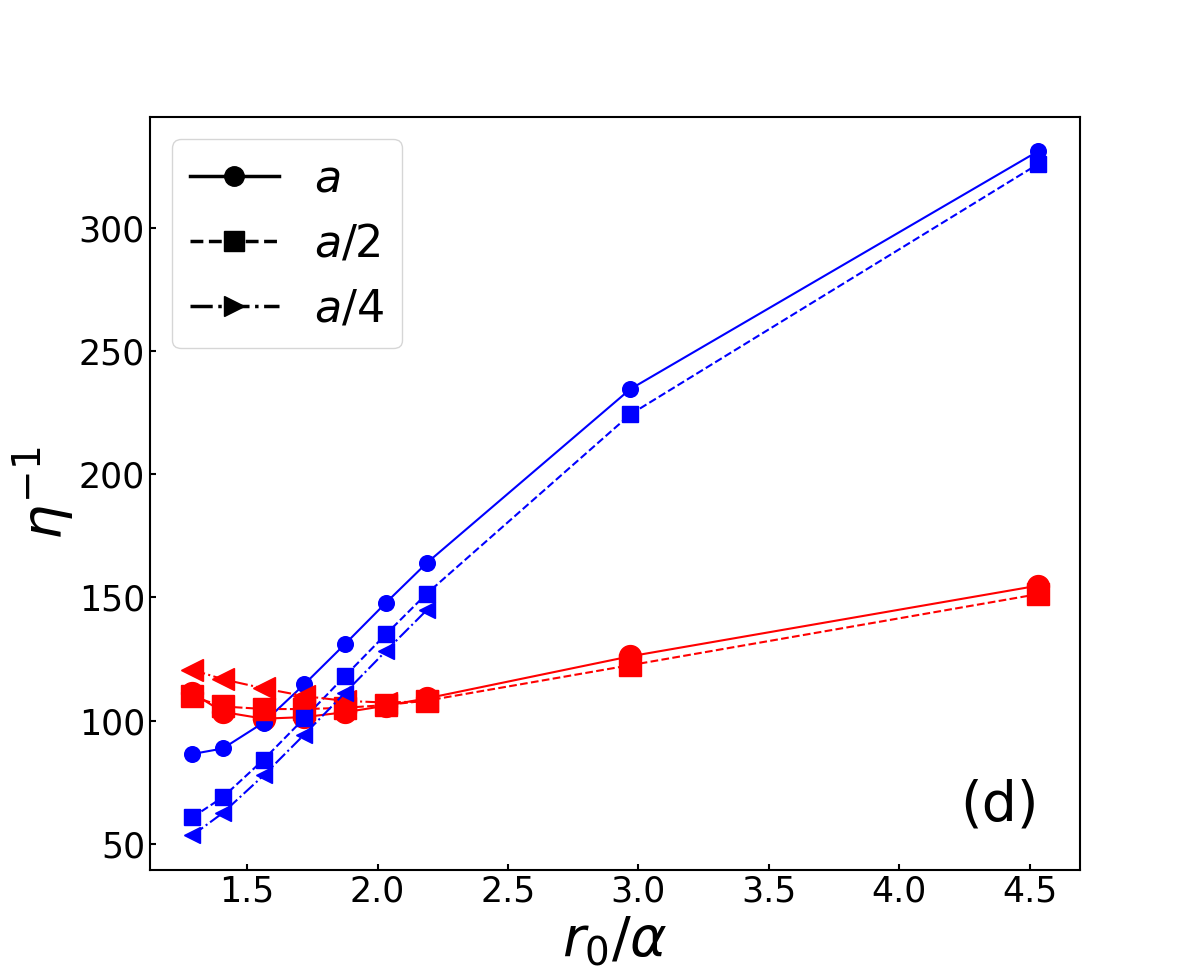}
    \caption{
      (a) Swimming speed, (b) flagellum angular speed,  (c) swimming power and (d) inverse of the efficiency vs.\ pipe radius for three different resolutions
      with blob radius $a$, $a/2$ and $a/4$.
      Red curve-symbols extended mode and blue curve-symbols wrapped mode.
    }
    \label{fig:resolution}
\end{figure}

Here we conduct a resolution study to verify the robustness of our results.
We reduce the blob radius used to discretize the bacterium body and the pipes from $a$ to $a/2$ and $a/4$
while at the same time increasing the resolution so the number of blobs are $\sim 4$ and $\sim 16$ larger than in the original model respectively.
We do not refine the flagellum as its thickness, proportional to their blobs radius, should remain constant during the resolution study.

For these three resolutions we measure the swimming speed ($u_z$),
flagellum rotational speed ($\omega_z$), swimming power ($P$) and inverse efficiency ($\eta^{-1}$).
The results are presented in Fig.\ \ref{fig:resolution}.
There are some differences in the swimming speed, see Fig.\ \ref{fig:resolution}a.
However, the other quantities only show a small deviation respect the original resolution.
The crossover in the inverse of the efficiency between the wrapped and the extended mode is more pronounced
for the highest resolution considered, see Fig.\ \ref{fig:resolution}d.
More important, the overall trends observed for all the magnitudes measured remain unchanged for all resolutions,
and thus, the results are robust against the resolution of the model.

%% \begin{figure}
%%     \centering
%%     \includegraphics[scale=0.4]{Supply_Figs/Shum_validation_aspect_ratio_height.png}
    
%%     \caption{
%%       Steady state distance above a no-slip surface during the circular trajectory. Here $\gamma$ is the aspect ratio and $\overline{c}$ is the radius of equivalent sphere.
%%     }
%%     \label{fig:height_profile}
%% \end{figure}

\bibliographystyle{jfm}
\bibliography{biblio.bib}

\end{document}